\documentclass{ws-ijmpcs}

\begin{document}

\markboth{Felix Wick}
{Results on Charm Baryon Spectroscopy from Tevatron}

%
\catchline{}{}{}{}{}
%

\title{RESULTS ON CHARM BARYON SPECTROSCOPY FROM TEVATRON}

\author{FELIX WICK\footnote{On behalf of the CDF Collaboration.}}

\address{Institut fuer Experimentelle Kernphysik, Karlsruhe Institute of Technology\\
Wolfgang-Gaede-Str. 1, Karlsruhe, 76131, Germany\\
wick@fnal.gov}

\maketitle


\begin{abstract}
  Due to an excellent mass resolution and a large amount of available
  data, the CDF experiment, located at the Tevatron proton-antiproton
  accelerator, allows the precise measurement of spectroscopic
  properties, like mass and decay width, of a variety of states. This
  was exploited to examine the first orbital excitations of the
  $\Lambda_c$ baryon, the resonances $\Lambda_c(2595)$ and
  $\Lambda_c(2625)$, in the decay channel $\Lambda_c^+\,\pi^+\,\pi^-$,
  as well as the $\Lambda_c$ spin excitations $\Sigma_c(2455)$ and
  $\Sigma_c(2520)$ in its decays to $\Lambda_c^+\,\pi^-$ and
  $\Lambda_c^+\,\pi^+$ final states in a data sample corresponding to
  an integrated luminosity of $5.2\,\mathrm{fb}^{-1}$. We present
  measurements of the mass differences with respect to the $\Lambda_c$
  and the decay widths of these states, using significantly higher
  statistics than previous experiments.
\end{abstract}

\ccode{PACS numbers: 14.20.Lq, 14.20.Gk}

\section{Introduction}

Heavy quark baryons provide an interesting laboratory for studying and
testing Quantum Chromodynamics (QCD), the theory of strong
interactions. The heavy quark states test regions of the QCD where
perturbation calculations cannot be used, and many different
approaches to solve the theory were developed. In this write-up we
focus on $\Lambda_c(2595)$, $\Lambda_c(2625)$, $\Sigma_c(2455)$ and
$\Sigma_c(2520)$ baryons. All four states were observed before and
some of their properties measured.\cite{PDG} However, different mass
measurements for $\Sigma_c(2520)$ and $\Lambda_c(2625)$ are
inconsistent and the statistics for $\Lambda_c(2595)$ and
$\Lambda_c(2625)$ is rather low. In addition, Blechman and co-workers
showed that a more sophisticated treatment, which would take into
account the proximity of the threshold in the $\Lambda_c(2595)$ decay,
yields a $\Lambda_c(2595)$ mass which is $2-3\,\mathrm{MeV}/c^2$ lower
than the one observed.\cite{Blechman} $\Sigma_c$ states were observed
and studied in $\Lambda_c\pi$ decays, while excited $\Lambda_c$ states
decay mainly to a $\Lambda_c\pi\pi$ final state and decays through
intermediated $\Sigma_c$ resonances are possible. One peculiarity of
the experimental studies of these baryons is in their cross talks,
which requires special care in the treatment of the background due to
different kinematic regions allowed for different sources.

In this analysis we exploit a large sample of $\Lambda_c^+\rightarrow
pK^-\pi^+$ decays collected by the CDF detector to perform the
measurement of the masses and widths of the discussed charmed baryons.
We take into account all cross-talks and threshold effects expected in
the decays under study.\cite{PublicNote} Throughout this document the
use of a specific particle state implies the use of the
charge-conjugate state as well.

\section{Analysis Strategy}

The employed dataset corresponds to an integrated luminosity of
$5.2\,\mathrm{fb^{-1}}$ and was collected using the displaced track
trigger which requires two charged particles coming from a single
vertex with transverse momenta larger than $2\,\mathrm{GeV}/c$ and
impact parameters in the plane transverse to the beamline in the
region of $0.1$ to $1\,\mathrm{mm}$.

The selection of the candidates is done in two steps. As all final
states feature a $\Lambda_c$ daughter, as first step we perform a
$\Lambda_c$ selection. In the second step we perform a dedicated
selection of the four states under study. In each step we use neural
networks to distinguish signal from background. All neural
networks\cite{Neurobayes} are trained using data only by means of the
$_s\mathcal{P}lot$ technique\cite{sPlot}. As we use only data for the
neural network trainings, for each case we split the sample to two
parts (even and odd event numbers) and train two networks. Each of
them is applied to the complementary subsample in order to maintain a
selection which is trained on a sample independent from the one to
which we apply it.

In order to determine the mass differences relative to the $\Lambda_c$
and the decay widths of the six studied states, we perform binned
maximum likelihood fits of three separate mass difference ($\Delta M$)
distributions. The first two are $\Lambda_c^+\pi^+$ and
$\Lambda_c^+\pi^-$ where the states $\Sigma_c(2455)^{++,0}$ and
$\Sigma_c(2520)^{++,0}$ are studied. The last one is
$\Lambda_c^+\pi^+\pi^-$ for $\Lambda_c(2595)^+$ and
$\Lambda_c(2625)^+$. In each of the distributions we need to
parametrize two signals and several background components.

\begin{figure}
\centering
a)
\includegraphics[width=.42\textwidth]{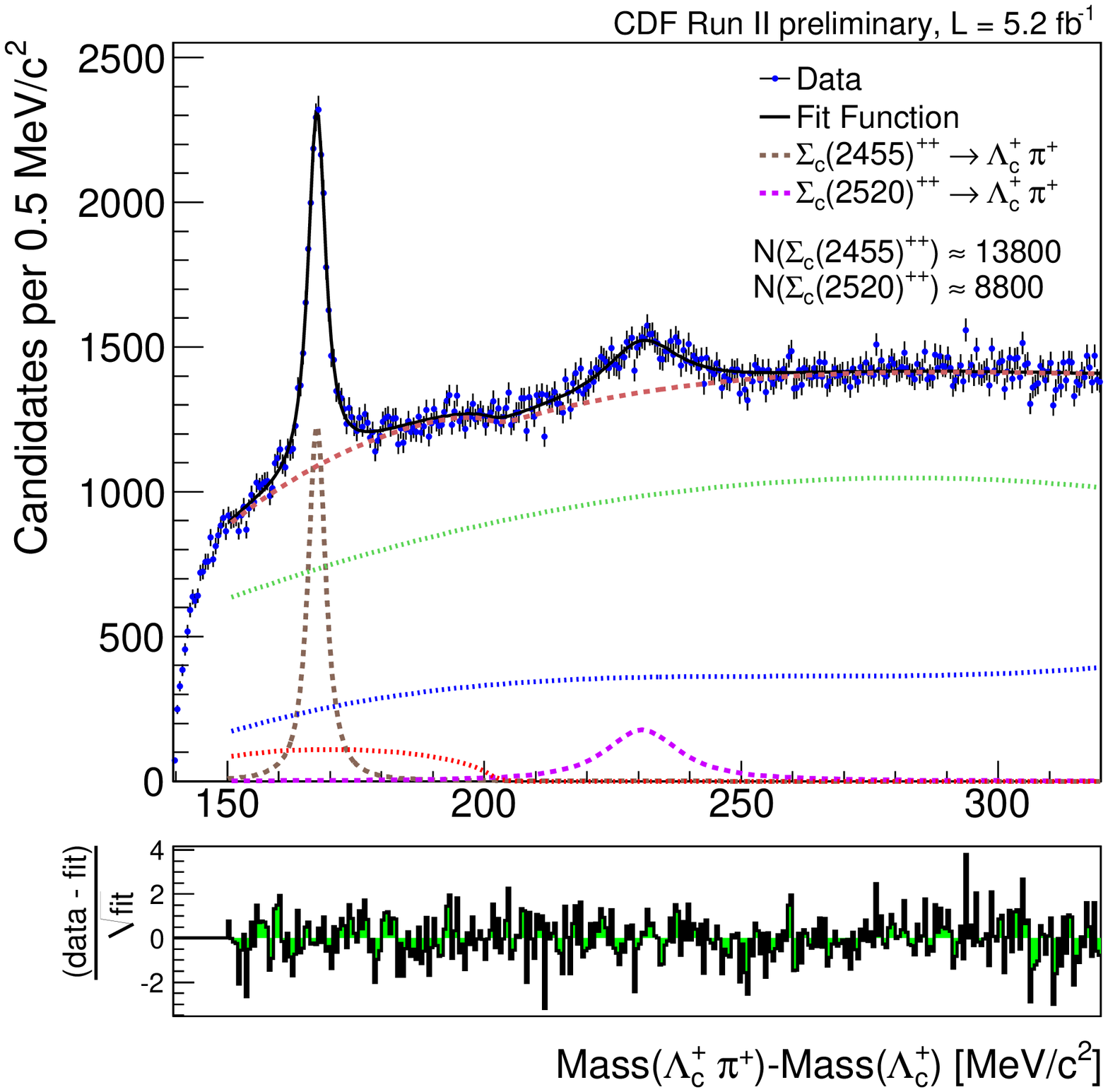}
\hspace{0.05\textwidth}
b)
\includegraphics[width=.42\textwidth]{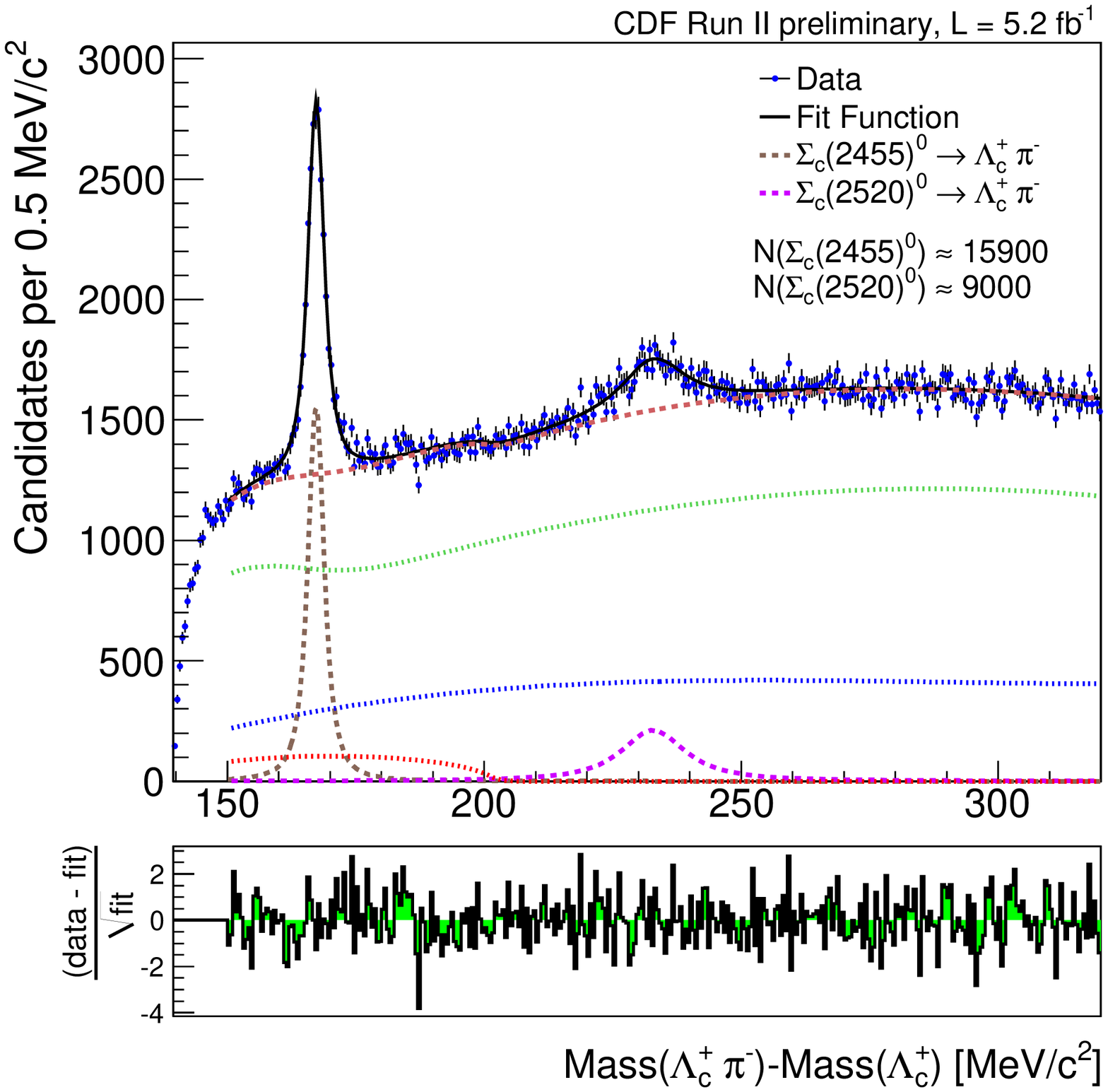}
\caption{The (a) $M(\Lambda_c^+\pi^+)-M(\Lambda_c^+)$ and (b)
  $M(\Lambda_c^+\pi^-)-M(\Lambda_c^+)$ distribution obtained from data
  (points with error bars) together with the fit projection (black
  line). The brown and purple lines correspond to the two signal
  contributions, the green line represents the combinatorial
  background without real $\Lambda_c$, the blue line shows real
  $\Lambda_c$ combined with a random pion and the red dotted line
  represents a reflection from excited $\Lambda_c$ decays. The red
  dashed line corresponds to the sum of all three background
  contributions.}
\label{fig:Sc_Fit}
\end{figure}

Both $\Sigma_c(2455)^{++,0}$ and $\Sigma_c(2520)^{++,0}$ are described
by a nonrelativistic Breit-Wigner function convolved with a resolution
function. The resolution function is parametrized with three Gaussians
centered in zero and other parameters derived from simulated events.
The mean width of the resolution function is about
$1.6\,\mathrm{MeV}/c^2$ for $\Sigma_c(2455)^{++,0}$ and about
$2.6\,\mathrm{MeV}/c^2$ for $\Sigma_c(2520)^{++,0}$. We consider three
different types of background, namely random combinations without real
$\Lambda_c$, combinations of real $\Lambda_c$ with a random pion and
events due to the decay of excited $\Lambda_c$ to
$\Lambda_c^+\pi^+\pi^-$. Thereby, random combinations without a real
$\Lambda_c$ dominate and are described by a second-order polynomial
with shape and normalization derived in a fit to the $\Delta M$
distribution from $\Lambda_c$ mass sidebands. The apparent difference
between doubly-charged and neutral combinations is due to
$D^*(2010)^+$ mesons with multibody $D^0$ decays, where not all $D^0$
decay products are reconstructed. In order to describe this
reflection, an additional Gaussian function is used. The second
background source, consisting of real $\Lambda_c$ combined with a
random pion, is modeled by a third-order polynomial. As we do not have
an independent proxy for this source, all parameters are left free in
the fit. The last source, originating from nonresonant
$\Lambda_c(2625)$ decays, is described by the projections of the flat
$\Lambda_c^+\pi^+\pi^-$ Dalitz plot on the appropriate
$\Lambda_c^+\pi^+$ respective $\Lambda_c^+\pi^-$ axes. The measured
data distributions together with the fit projections can be found in
figure \ref{fig:Sc_Fit}.

\begin{figure}
\centering
\includegraphics[width=.42\textwidth]{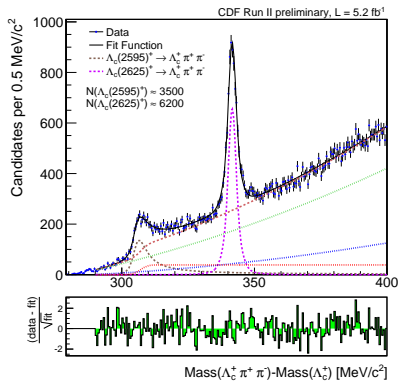}
\caption{The $M(\Lambda_c^+\pi^+\pi^-)-M(\Lambda_c^+)$ distribution
  obtained from data (points with error bars) together with the fit
  projection (black line). The brown and purple lines correspond to
  the two signal contributions, the green line represents the
  combinatorial background without real $\Lambda_c$, the blue line
  shows real $\Lambda_c$ combined with two random pions and the red
  dotted line represents real $\Sigma_c(2455)$ combined with a random
  pion.  The red dashed line corresponds to the sum of all three
  background contributions.}
\label{fig:LcS_Fit}
\end{figure}

In the fit for $\Lambda_c(2595)$ and $\Lambda_c(2625)$, an additional
complication compared to the $\Sigma_c$ case arises from the fact that
previous measurements of the $\Lambda_c(2595)$ properties indicate
that it decays dominantly to $\Sigma_c(2455)\pi$, which has its
threshold very close to the $\Lambda_c(2595)$ mass.
Ref.~\refcite{Blechman} shows that taking into account the resulting
strong variation of the natural width yields a lower $\Lambda_c(2595)$
mass than observed by former experiments. With CDF statistics we are
much more sensitive to the details of the $\Lambda_c(2595)$ line shape
than previous analyses. The $\Lambda_c(2595)$ parametrization follows
Ref.~\refcite{Blechman}. The state is described by a nonrelativistic
Breit-Wigner function with a mass-dependent decay width which is
calculated as integral over the Dalitz plot containing the
$\Sigma_c(2455)$ resonances. The pion coupling constant $h_2$ is
related to the $\Lambda_c(2595)$ decay width and represents the
quantity we measure instead of the natural width. The shape is then
numerically convolved with a three Gaussian resolution function
determined from simulation, which has a mean width of about
$1.8\,\mathrm{MeV}/c^2$. The signal function for the $\Lambda_c(2625)$
is a nonrelativistic Breit-Wigner which is again convolved with a
three Gaussian resolution function determined from simulation. The
mean width of the resolution function is about
$2.4\,\mathrm{MeV}/c^2$. The background consists of three different
sources which are combinatorial background without real $\Lambda_c$,
real $\Lambda_c$ combined with two random pions and real
$\Sigma_c^{++,0}$ combined with a random pion. The combinatorial
background without real $\Lambda_c$ is parametrized by a second order
polynomial whose parameters are determined in a fit to the $\Delta M$
distribution of candidates from $\Lambda_c$ mass sidebands. The second
source, consisting of real $\Lambda_c$ combined with two random pions,
is parametrized by a second order polynomial with all parameters
allowed to float in the fit. The model of the final background source,
real $\Sigma_c(2455)$ combined with a random pion, is based on a
uniform function defined from the threshold to the end of the fit
range, where the natural widths as well as the resolution effects of
the $\Sigma_c(2455)^{++,0}$ are taken into account for the threshold
determination. The size of this contribution is constrained to the
$\Sigma_c(2455)$ yield obtained from fits to the projections of the
$\Lambda_c^+\pi^+\pi^-$ Dalitz plot from the upper $\Lambda_c(2625)$
mass sideband on the appropriate $\Lambda_c^+\pi^+$ respective
$\Lambda_c^+\pi^-$ axes. The measured data distribution together with
the fit projection can be found in figure \ref{fig:LcS_Fit}.

We investigate several systematic effects which can affect our
measurements. Generally, they can be categorized as imperfect modeling
by the simulation, imperfect knowledge of the momentum scale of the
detector, ambiguities in the fit model and uncertainties on the
external inputs to the fit of the $\Lambda_c(2595)$ signal. The most
critical point is that we need to understand the intrinsic resolution
of the detector in order to properly describe the signal shapes,
because an uncertainty in the resolution has a large impact on the
determination of the natural widths. As we estimate it using simulated
events, it is necessary to substantiate that the resolution obtained
from simulation agrees with the one in real data. We use $D^*(2010)^+
\rightarrow D^0 \pi^+$ with $D^0 \rightarrow K^- \pi^+$ and $\psi(2S)
\rightarrow J/\psi \pi^+ \pi^-$ with $J/\psi \rightarrow \mu^+ \mu^-$
for this purpose and compare the resolution in data and simulated
events as a function of the transverse momentum of the pions added to
$D^0$ or $J/\psi$. We also compare the overall resolution scale
between data and simulated events and find that all variations are
below 20\% which we assign as uncertainty on our knowledge of the
resolution function.

\section{Results}

We select about 13800 $\Sigma_c(2455)^{++}$, 15900
$\Sigma_c(2455)^{0}$, 8800 $\Sigma_c(2520)^{++}$, 9000
$\Sigma_c(2520)^{0}$, 3500 $\Lambda_c(2595)^+$ and 6200
$\Lambda_c(2625)^+$ signal events and obtain the following results for
the mass differences relative to the $\Lambda_c$ mass and the decay
widths:
\begin{eqnarray}
  \Delta M(\Sigma_c(2455)^{++})&=&167.44 \pm 0.04\,\mathrm{(stat.)} \pm 0.12\,\mathrm{(syst.)}\, \mathrm{MeV}/c^2, \nonumber \\
  \Gamma(\Sigma_c(2455)^{++})&=&2.34 \pm 0.13\,\mathrm{(stat.)} \pm 0.45\,\mathrm{(syst.)}\, \mathrm{MeV}/c^2, \nonumber \\
  \Delta M(\Sigma_c(2455)^{0})&=&167.28 \pm 0.03\,\mathrm{(stat.)} \pm 0.12\,\mathrm{(syst.)}\, \mathrm{MeV}/c^2, \nonumber \\
  \Gamma(\Sigma_c(2455)^{0})&=&1.65 \pm 0.11\,\mathrm{(stat.)} \pm 0.49\,\mathrm{(syst.)}\, \mathrm{MeV}/c^2, \nonumber \\
  \Delta M(\Sigma_c(2520)^{++})&=&230.73 \pm 0.56\,\mathrm{(stat.)} \pm 0.16\,\mathrm{(syst.)}\, \mathrm{MeV}/c^2, \nonumber \\
  \Gamma(\Sigma_c(2520)^{++})&=&15.03 \pm 2.12\,\mathrm{(stat.)} \pm 1.36\,\mathrm{(syst.)}\, \mathrm{MeV}/c^2, \nonumber \\
  \Delta M(\Sigma_c(2520)^{0})&=&232.88 \pm 0.43\,\mathrm{(stat.)} \pm 0.16\,\mathrm{(syst.)}\, \mathrm{MeV}/c^2, \nonumber \\
  \Gamma(\Sigma_c(2520)^{0})&=&12.51 \pm 1.82\,\mathrm{(stat.)} \pm 1.37\,\mathrm{(syst.)}\, \mathrm{MeV}/c^2, \nonumber \\
  \Delta M(\Lambda_c(2595)^+)&=&305.79 \pm 0.14\,\mathrm{(stat.)} \pm 0.20\,\mathrm{(syst.)}\, \mathrm{MeV}/c^2, \nonumber \\
  h_2^2(\Lambda_c(2595)^+)&=&0.36 \pm 0.04\,\mathrm{(stat.)} \pm 0.07\,\mathrm{(syst.)},\nonumber \\
  \Delta M(\Lambda_c(2625)^+)&=&341.65 \pm 0.04\,\mathrm{(stat.)} \pm 0.12\,\mathrm{(syst.)}\, \mathrm{MeV}/c^2. \nonumber 
\end{eqnarray}
For the width of the $\Lambda_c(2625)$ we get a value consistent with
zero and therefore calculate an upper limit using a Bayesian approach
with a uniform prior restricted to positive values. At the 90\% C.L.
we obtain $\Gamma(\Lambda_c(2625)^+)<0.97\,\mathrm{MeV}/c^2$. For
easier comparison to previous results, $h_2^2$ corresponds to
$\Gamma(\Lambda_c(2595)^+)=2.59 \pm 0.30 \pm 0.47\,\mathrm{MeV}/c^2$.

Except of $\Delta M(\Lambda_c(2595)^+)$, all measured quantities are
in agreement with previous measurements. For $\Delta
M(\Lambda_c(2595)^+)$ we observe a value which is by
$3.1\,\mathrm{MeV}/c^2$ smaller than the existing world average. This
difference is of the same size as found in Ref.~\refcite{Blechman}. The
precision for the $\Sigma_c$ states is comparable to the precision of
the world averages. In the case of excited $\Lambda_c$ states, our
results provide a significant improvement in precision against
previous measurements.

\end{document}